\documentclass[aps,12pt,preprintnumbers,showpacs]{revtex4}
\usepackage{graphicx}
\usepackage{epstopdf}
\usepackage{amsmath,amssymb,bm}
\renewcommand\d{\partial}

\renewcommand\r{{\bf r}}

\begin{document}
\preprint{INT-PUB 07-17}
\title{Coulomb impurity in graphene}
\author{Rudro R. Biswas}
\affiliation{Department of Physics, Harvard University, Cambridge MA
02138, USA}

\author{Subir Sachdev}
\affiliation{Department of Physics, Harvard University, Cambridge MA
02138, USA}

\author{Dam~T.~Son}
\affiliation{Institute for Nuclear Theory, University of Washington,
Seattle, Washington 98195, USA}
\date{June 2007}
\begin{abstract}
\noindent
We consider the problem of screening of an electrically charged impurity 
in a clean graphene sheet.  When electron-electron interactions are neglected, the screening charge has a sign opposite to that of the impurity, and is localized near the impurity.
Interactions between electrons smear out the induced charge density
to give a large-distance tail that follows approximately, 
but not exactly, an $r^{-2}$ behavior and with a sign which is the
{\em same\/} as that of the impurity. 

\end{abstract}
\maketitle

\section{Introduction.}

With the recent explosion of interest in graphene, there are numerous
experimental motivations for understanding the influence of impurities on its
electronic and transport properties. For non-interacting electrons, the influence
of a dilute concentration of impurities on transport properties 
has been investigated in some depth \cite{ostrovsky}.
Here we shall instead study in some detail the physics associated with a {\em single\/} impurity
carrying electrical charge $Z$. 
Nanoscale studies of the electronic properties 
of a single graphene sheet have recently become possible \cite{yacoby,stm}, and so it 
should eventually be possible to observe the variation in the charge density and the
local density of states as a function of distance from the impurity.
We shall show here that this spatial structure is a sensitive probe of the strong correlations
between the electrons in graphene, and of the unusual nature of screening in a two-dimensional semi-metal
with a Dirac dispersion spectrum.

For non-interacting electrons, the influence of a Coulomb impurity exerting a potential
$Z e^2 /(4 \pi \epsilon_0 r)$ (where $r$ is the distance from the impurity) was studied some time
ago \cite{vm}. This case is equivalent to the familiar ``Friedel problem'' but for Dirac fermions.
However, even for this seemingly simple case, there are subtleties which were overlooked
in the initial treatment \cite{vm}, and corrected in Ref.~\onlinecite{qimp}.
A number of papers appeared \cite{shytov,novikov,castroneto} while our paper was being written,
presenting additional results on this non-interacting problem. 
We shall review and extend the results of Ref.~\onlinecite{qimp} for non-interacting electrons in Section~\ref{sec:nonint}.
We shall then proceed to the full treatment of the impurity problem, and allow for electron-electron Coulomb interactions. 
%As we shall see in Section~\ref{sec:int}, this leads to dramatic changes in the nature
%of the results, and the spatial dependence of the screening charge becomes a sensitive probe of ``quantum critical''

In short, our results are as follows.  For noninteracting electrons, 
the screening charge is a local delta-function in space to all orders in
perturbation theory over the impurity charge.  The sign of this screening
charge is opposite to that of the impurity, as is usually the case.
However, once interaction between electrons is turned on, 
the screening charge develops a long-range tail, 
even for small impurity charges. 
The tail follows approximately an $r^{-2}$ law, 
with a coefficient which varies quite slowly with $r$. Notably, the sign of this
tail is the same as that of the impurity. 
The long-range tail of the screening charge, thus, is a sensitive probe
of the interaction between electrons, 
in particular to the renormalization of the fermion velocity and the ``quantum critical'' 
aspects \cite{son} of the interacting Dirac fermion problem.

Let us begin with a statement of the problem. After taking the continuum limit to $N=4$ species
of two-component Dirac fermions $\Psi_a$ ($a = 1 \ldots N$) we have the theory defined
by the Euclidean partition function
\begin{eqnarray}
\mathcal{Z} &=& \int \mathcal{D} \Psi_\alpha\, \mathcal{D} A_\tau\, 
\exp \left( - \mathcal{S} - \mathcal{S}_{\rm imp} \right), \nonumber \\
\mathcal{S} &=& \sum_{a=1}^N \int d^2 r \int d \tau\,  \Psi_a^\dagger ({\bf r}, \tau) \left[ \frac{\partial}{\partial\tau} + 
i A_\tau ({\bf r}, \tau) + i v \sigma^x \frac{\partial}{\partial x}
+ i v \sigma^y \frac{\partial}{\partial y} \right] \Psi_a \nonumber ({\bf r}, \tau) \\
&~&~~~~~~~~~~~~~~ + \frac{1}{2g^2} \int \frac{d^2 q}{4 \pi^2} \int d\tau\, 2 q \left| A_\tau ({\bf q},\tau) \right|^2, \nonumber\\
\mathcal{S}_{\rm imp} &=& -i Z \int d\tau A_\tau ({\bf r}=0,\tau).
\label{zz}
\end{eqnarray}
The functional integral is over fields defined in two 
spatial dimensions ${\bf r}=(x,y)$ and imaginary time $\tau$, $\sigma^{x,y}$ are
Pauli matrices acting on the Dirac space, and $v$ is the Fermi velocity. 
The scalar potential which mediates the $e^2/(4 \pi \epsilon_0 |\r|)$ Coulomb interaction between the electrons
is $i A_\tau ({\bf r}, \tau)$; after a spatial Fourier transform to two-dimensional momenta ${\bf q}$, this interaction requires
the $2q$ ($=2|{\bf q}|$) co-efficient of the term quadratic in $A_\tau$, with the coupling $g^2 = e^2/\epsilon_0$.
The screening due to a substrate of dielectric constant $\varepsilon$ can also be included by modifying the
coupling to \cite{son} $g^2 = 2e^2 /(\epsilon_0 (1+ \varepsilon))$. The action $\mathcal{S}$ therefore represents the
physics of an ideal graphene layer. 
The influence of an impurity of net charge $Z$ at ${\bf  r}=0$ is described by $\mathcal{S}_{\rm imp}$.

Many essential aspects of the theory above follow from its properties under the renormalization group (RG) transformation under
which ${\bf r } \rightarrow {\bf r}/s$ and $\tau \rightarrow \tau/s$. 
A standard analysis shows that all three couplings in $\mathcal{Z}$, namely $v$, $Z$, and $g$, 
are invariant under this transformation at tree level.  Indeed, for two of the couplings, this invariance extends
to all orders in perturbation theory: the coupling $g$ does not renormalize because of the non-analytic $q$ co-efficient,
while $Z$ remains invariant because it is protected by gauge invariance \cite{qimp}. 
So we need only examine the RG flow of a single
coupling, the velocity $v$. Because $v$ is a bulk coupling, its flow cannot be influenced in the thermodynamic limit
by a single impurity, and so can be computed in the absence of the impurity. Such a RG flow was initially examined in the more
general context of theories with Chern-Simons couplings in Ref.~\onlinecite{ys}, but a complete presentation was given
in the present context in Ref.~\onlinecite{son}: we shall use the notation and results of the latter paper here,
with the exception that we use two-component Dirac fermions with
$N=4$ while Ref.~\onlinecite{son} uses four-component Dirac fermions
with $N=2$.

It will be useful for our analysis to introduce two combinations of the above couplings which also have engineering
dimension zero, and hence are pure numbers.  These are
\begin{equation}
\lambda = \frac{g^2 N}{32 \hbar v}~~~~;~~~~\alpha = \frac{g^2 Z}{4 \pi \hbar v}
\label{pars}
\end{equation}
(we have set $\hbar=1$ elsewhere in the paper).
As we will see, the coupling $\lambda$ is a measure of the strength of the electron-electron Coulomb interactions, 
while $\alpha$ measures the strength of the electron-impurity Coulomb interaction.

We shall limit our explicit results here to the spatial form of the charge density
\begin{equation}
n (r) = -\sum_a \mbox{Tr} \langle \Psi^\dagger_a ({\bf r}, \tau) \Psi_a ({\bf r},\tau) \rangle, \label{nr}
\end{equation}
(where $\mbox{Tr}$ acts on the Dirac space) induced by the impurity. However, our RG strategy can be extended
to other observables of experimental interest, such as the local density of states.

As noted above, we will begin in Section~\ref{sec:nonint} by considering only the 
electron-impurity Coulomb interaction, while electron-electron Coulomb interactions will be accounted for in
Section~\ref{sec:int}.

\section{Non-interacting electrons}
\label{sec:nonint}

This section will ignore the electron-electron Coulomb interactions. 
Formally, we work in the limit $\lambda\to0$, but $\alpha$ is kept fixed.
The problem reduces to that of a single
Dirac electron in the attractive impurity potential
\begin{equation}
V(r) = - \frac{Zg^2}{4 \pi r}. \label{vr}
\end{equation}
This problem was originally studied in Ref.~\onlinecite{vm}. 
However, they introduced an arbitrary cutoff at high energy to regulate the 
problem at short distances, and this leads
to spurious results \cite{qimp}. As we will demonstrate here, there is no dependence upon
a cutoff energy scale at all orders in perturbation theory, provided the high energy behavior
is regulated in a proper gauge-invariant manner. With no cutoff energy scale
present, a number of results can be deduced by simple dimensional analysis. The Fourier transform
of the charge density $n(r)$ is dimensionless, and therefore we can write
\begin{equation}
n(q) = - N F(\alpha), \label{n1}
\end{equation}
where $F(\alpha)$ is a universal function of the dimensionless coupling $\alpha$. Note that $n(q)$
is required by this dimensional argument to be $q$-independent, and 
so $n(r) \propto \delta^2 ({\bf r})$. 

The arguments so far are perturbative, but non-perturbative effects can be deduced
by solving the full Dirac equation in the potential in Eq.~(\ref{vr}). This solution has appeared
elsewhere \cite{shytov,novikov,castroneto}, and so we will not reproduce it here. 
Such an analysis shows that the perturbative arguments apply for $\alpha < 1/2$, but
new physics appears for $\alpha > 1/2$. In particular, Shytov {\em et al.} \cite{shytov}
showed that $n(r) \sim -r^{-2}$ for $\alpha > 1/2$ (the sign of this tail is opposite to
that of the impurity).

We shall limit our discussion in this section to the $\alpha<1/2$ case. One reason for doing so is that
electron-electron Coulomb interactions act to reduce the effective value of $\alpha$.
This will become clearer in Section~\ref{sec:int}, but we note here that a 
standard RPA screening of the potential $V(r)$ in Eq.~(\ref{vr})
can be simply accounted for by applying the mapping
\begin{equation}
\alpha \rightarrow \frac{\alpha}{1+ \lambda}
\end{equation}
to the results of the present section. The value of $\lambda$ in graphene is not small \cite{son}.

\begin{figure}[t]
\centering \includegraphics[width=3.5in]{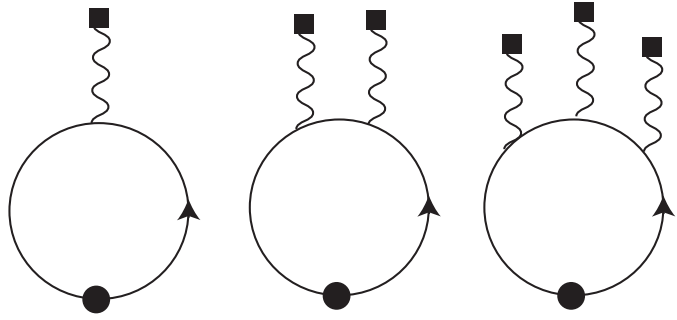}
\caption{Feynman diagrams for the charge density without electron-electron interactions
to order $\alpha^3$.
The filled square is the impurity site, the wavy line is the $A_\tau$ propagator,
the line is the fermion propagator, and the filled circle is the charge density operator.
}
\label{diags}
\end{figure}

We shall now establish the existence of the universal function $F(\alpha)$ in Eq.~(\ref{n1})
to all orders in $\alpha$. 
The existence of a universal $F(\alpha)$ is a consequence of the
non-renormalization of the impurity charge $Z$ \cite{qimp}.
We compute $n(q)$ diagrammatically, and the needed diagrams
all have one fermion loop and are shown in Fig.~\ref{diags}. 

To first order in $\alpha$ we have
\begin{equation}
n(q) = -\frac{Z}{2q} \Pi_0 (q),
\label{p0}
\end{equation}
where $\Pi_0 (q)$ is the bare polarization operator
\begin{eqnarray}
\Pi_0 (q) &=& - g^2 N \int \frac{d^2 k}{4 \pi^2} \int \frac{d \omega}{2\pi}
\mbox{Tr} \left[ \left( - i \omega +
v {\bf k} \cdot \vec{\sigma} \right)^{-1} \left( - i \omega + v ({\bf k}
+ {\bf q}) \cdot \vec{\sigma} \right)^{-1} \right] \nonumber \\
&=& \frac{g^2 N q}{16 v}\,, \label{p0res}
\end{eqnarray}
and so we have $F(\alpha) = (\pi/8) \alpha + \mathcal{O} (\alpha^2)$. 

The order $\alpha^2$ graph in Fig.~\ref{diags} vanishes by Furry's theorem,
and at order $\alpha^3$ we write the contribution to $n(q)$ in the form
\begin{equation}  N (Z g^2)^3 \int \frac{d^2
k_1}{4 \pi^2} \frac{d^2 k_2}{4 \pi^2} \frac{d^2 k_3}{4 \pi^2}
\frac{A({\bf k}_1, {\bf k}_2 , {\bf k}_3)}{8 k_1 k_2 k_3} (2 \pi)^2
\delta^2 ({\bf k}_1 + {\bf k}_2 + {\bf k}_3 + {\bf q}), \label{rho3}
\end{equation}
where
\begin{eqnarray}
&& A({\bf k}_1, {\bf k}_2 , {\bf k}_3) = \int \frac{d^2 p}{4 \pi^2}
\int \frac{d \omega}{2 \pi} \mbox{Tr} \left[ \left( - i \omega +
v{\bf p} \cdot \vec{\sigma} \right)^{-1} \left( - i \omega + v({\bf p}
+ {\bf q}) \cdot \vec{\sigma} \right)^{-1} \right. \nonumber \\
&&~~~~~~~~~~~~~\times \left. \left( - i \omega + v({\bf
p} + {\bf q} + {\bf k}_1) \cdot \vec{\sigma} \right)^{-1} \left(
- i \omega + v({\bf p} + {\bf q}+ {\bf k}_1 + {\bf k}_2) \cdot
\vec{\sigma} \right)^{-1} \right],
\end{eqnarray}
where it is understood here and below that $-{\bf q} = {\bf k}_1 +
{\bf k}_2 + {\bf k}_3$. We now want to symmetrize this by placing
the external vertex with momentum ${\bf q}$ at different points on
the loop --- this should not change the final result for $n(q)$.
In this manner we obtain
\begin{eqnarray}
&& 3 A({\bf k}_1, {\bf k}_2 , {\bf k}_3) = \int \frac{d^2 p}{4 \pi^2}
\int \frac{d \omega}{2 \pi} \Biggl\{ \mbox{Tr} \left[ \left( - i \omega +
v{\bf p} \cdot \vec{\sigma} \right)^{-1} \left( - i \omega +v ({\bf p}
+ {\bf q}) \cdot \vec{\sigma} \right)^{-1} \right. \nonumber \\
&&~~~~~~~~~~~~~\times \left. \left( - i \omega + v ({\bf
p} + {\bf q} + {\bf k}_1) \cdot \vec{\sigma} \right)^{-1} \left(
- i \omega + v ({\bf p} + {\bf q}+ {\bf k}_1 + {\bf k}_2) \cdot
\vec{\sigma} \right)^{-1} \right] \nonumber \\
&&~~~~~~~~~~~~~~~~~~+\mbox{Tr} \left[ \left( - i \omega + v
{\bf p} \cdot \vec{\sigma} \right)^{-1} \left( - i \omega + v ({\bf p}
+ {\bf k}_1) \cdot \vec{\sigma} \right)^{-1} \right. \nonumber \\
&&~~~~~~~~~~~~~\times \left. \left( - i \omega + v ({\bf
p} + {\bf q} + {\bf k}_1) \cdot \vec{\sigma} \right)^{-1} \left(
- i \omega + v ({\bf p} + {\bf q}+ {\bf k}_1 + {\bf k}_2) \cdot
\vec{\sigma} \right)^{-1} \right] \nonumber \\
&&~~~~~~~~~~~~~~~~~~+\mbox{Tr} \left[ \left( - i \omega + v
{\bf p} \cdot \vec{\sigma} \right)^{-1} \left( - i \omega + v ({\bf p}
+ {\bf k}_1) \cdot \vec{\sigma} \right)^{-1} \right. \nonumber \\
&&~~~~~~~~~~~~~\times \left. \left( - i \omega + v ({\bf
p} + {\bf k}_1 + {\bf k}_2) \cdot \vec{\sigma} \right)^{-1} \left(
- i \omega + v ({\bf p} + {\bf q}+ {\bf k}_1 + {\bf k}_2) \cdot
\vec{\sigma} \right)^{-1} \right] \Biggr\}.
\end{eqnarray}
Now this expression has the important property that it vanishes at ${\bf
q}=0$, where we have
\begin{eqnarray}
&& 3 A({\bf k}_1, {\bf k}_2 , {\bf k}_3) = \int \frac{d^2 p}{4
\pi^2} \int \frac{d \omega}{2 \pi} \frac{\partial}{i\partial \omega}
\mbox{Tr} \left[ \left( - i \omega + v{\bf p} \cdot \vec{\sigma}
\right)^{-1} \left( - i \omega + v({\bf p}  + {\bf k}_1) \cdot
\vec{\sigma} \right)^{-1} \right. \nonumber
\\
&&~~~~~~~~~~~~~~~~~~~~~~~~~~~~~~~~~~~~~~~~~~~~~\times \left. \left(
- i \omega + v({\bf p} + {\bf k}_1 + {\bf k}_2) \cdot \vec{\sigma}
\right)^{-1} \right].
\end{eqnarray}
This property allows us to establish that the integral in
Eq.~(\ref{rho3}) is convergent and cut-off independent. Let the loop
momenta $p$, $k_1$, $k_2$, and $k_3$ all become much larger than the
external momentum $q$. The resulting integrand will scale as the
power of momenta associated with a logarithmic dependence on the
upper cutoff. However, in this limit of small $q$ we have just
established that the integrand is zero.
It is clear that this argument can be extended to all orders in $\alpha$. We have thus
established the existence of the cut-off independent function $F(\alpha)$.
We computed the integral in Eq.~(\ref{rho3}) numerically, and so obtained
\begin{equation}
F(\alpha) = \frac{\pi}{8} \alpha + (0.19 \pm 0.01) \alpha^3 + \mathcal{O} (\alpha^5).
\label{Falpha}
\end{equation}

\section{Interacting electrons}
\label{sec:int}

We will now consider the full problem defined in Eq.~(\ref{zz}), 
and account for both the electron-electron and electron-impurity 
Coulomb interactions. 

The problem can be solved in two limits: in the weak coupling limit
$\lambda\to0$ and the large $N$ limit, $N\to\infty$ with fixed $Z=O(1)$.
In both cases $\alpha/(1+\lambda)\ll1$, so one can
limit oneself to linear response in which the induced charge is 
[generalizing Eq.~(\ref{p0})]
\begin{equation}
  n(q) = - Z D(q)\Pi(q),
\end{equation}
where $D(q)$ is the full propagator of the Coulomb potential $A_\tau$,
and $\Pi(q)$ is the polarization tensor.  The connection between
$D(q)$ and $\Pi(q)$ is
\begin{equation}
  D^{-1}(q) = D_0^{-1}(q) + \Pi(q),
\end{equation}
where $D_0(q)$ is the bare propagator, 
\begin{equation}
  D_0(q) = \frac1{2q}\,.
\end{equation}
To leading order (either in coupling or $1/N$), the polarization
operator was given in Eq.~(\ref{p0res}), and we showed in Section~\ref{sec:nonint}
that this gives rise to a $q$-independent $n(q)$, or a
screening charge localized at $\r=0$.

However, if we compute corrections, we find logarithmically divergent
diagrams, where the logarithms are cut off from above by the inverse
lattice size and from below by $q$.  The leading logarithms are summed
by a standard RG procedure.  Since the theory is
renormalizable, we can eliminate the dependence on the cutoff by
expressing the each diagram in terms of the renormalized parameters,
instead of the bare parameters of the Lagrangian.  Choosing the
renormalization point to be $q_0$, and denote $v_0$ as the fermion
velocity at the scale $v$, the polarization tensor can be
schematically written as
\begin{equation}
  \Pi(q) = \Pi(q;q_0,v_0).
\end{equation}
In $\Pi$ there are logarithms of the ratio $q/q_0$.  We notice
that $\Pi(q;q_0,v_0)$ is is invariant under a change of the
renormalization $q_0$, given that $v_0$ is changed correspondingly
(the particle density has no anomalous dimension).
To eliminate the powers of $\log(q/q_0)$ we can choose $q_0=q$, hence
\begin{equation}
  \Pi(q) = \Pi(q;q,v(q)),
\end{equation}
where in the perturbative expansion of the right hand side there is no
large logarithms.  Thus to leading order it is given by a single diagram,
which was computed previously [Eq.~(\ref{p0res})],
\begin{equation}
  \Pi(q) = \frac{g^2N}{16 v(q)} q.
\end{equation}
All the leadings logarithms are contained in the function $v(q)$,
which satisfies the equation
\begin{equation}\label{RG-v}
  q\frac\d{\d q} v(q) = \beta(v),
\end{equation}
with the boundary condition $v(q_0)=v_0$.  The screening charge is
then
\begin{equation}\label{nlambda}
  n(q) = - Z \frac{\lambda(q)}{1+\lambda(q)}\,, 
  \qquad \lambda(q) = \frac{g^2N}{32v(q)}\,.
\end{equation}
The problem is now reduced to the problem of finding $v(q)$ [or,
equivalently, $\lambda(q)$].  This problem 
%was solved in
has a long history~\cite{Gonzalez}; most recently it has been
revisited in
Ref.~\onlinecite{son} (see also below).

To find the spatial charge distribution $n(\r)$ one needs to take
Fourier transform of Eq.~(\ref{nlambda}).  First one notice that if
the velocity does not run then $n(\r)$ is proportional to
$\delta(\r)$.  Only when $v$ runs with the momentum scale does $n(\r)$
differ from $0$ away from the origin.  When the running is slow (as at
weak coupling or at large $N$), the amount of screening charge
enclosed inside a circle of radius $r$ (assumed to be much larger than
the lattice spacing), to leading order, is
\begin{equation}\label{nr-int}
  \int\limits^r\! d\r'\, n(r') \approx n(q)|_{q=1/r} 
   = - Z \frac{\lambda(q)}{1+\lambda(q)}\biggl|_{q=1/r}.
\end{equation}
The total screening charge is small if $\lambda$ at the scale $1/r$ is
small, and close to $-1$ if $\lambda$ is large. Differentiating both
sides of Eq.~(\ref{nr-int}) with respect to $r$, one finds
\begin{equation}\label{nr-beta}
  n(r) = - \frac Z{2\pi r^2} \, \frac{\lambda(q)}{[1+\lambda(q)]^2}\,
  \frac{\beta(v(q))}{v(q)}\,.
\end{equation}
Note that the beta function for $v$ is negative, therefore we arrive
to a counterintuitive result the screening charge is \emph{positive}.
To see what is happening, let us take the limit $r\to\infty$ in
Eq.(\ref{nr-int}).  This limit corresponds to the infrared limit
$q\to0$.  We know that asymptotically $v(q)$ grows to $\infty$ in this
limit (although only logarithmically), hence
\begin{equation}
  \int\limits^\infty d\r'\, n(r') = 0.
\end{equation}
i.e., the total screening charge is zero when integrated over the
whole space (although the integral goes to zero very slowly).  The
presence of an external ion, therefore, only leads to charge
redistribution: a fraction of the unit charge is pushed from short
distance (of order of lattice spacing) to longer distances, but none
of the charge goes to infinity.  Therefore, there is a finite negative
screening charge localized near $\r=0$.  Its value can be found by
taking $r$ to be of order of inverse lattice spacing $a^{-1}$ in
Eq.~(\ref{nr-int}).  The final result for the screening charge density
can be written as
\begin{equation}\label{nr-2terms}
  n(r) = - Z \frac{\lambda(a^{-1})}{1+\lambda(a^{-1})} \delta(\r) 
  - \frac Z{2\pi r^2} \, \frac{\lambda(q)}{[1+\lambda(q)]^2}\,
  \frac{\beta(v(q))}{v(q)}\,.
\end{equation}
In the rest of the note we will concentrate our attention on the
long-distance tail of $n(r)$, ignoring the delta function at the
origin.

At weak coupling ($\lambda\ll1$), the beta function for $v(q)$ is
\begin{equation}
  \beta(v) = - \frac{g^2}{16\pi}\,.
\end{equation}
The solution to the RG equation, with the boundary condition $v=v_0$
at $q=q_0$, is
\begin{equation}
  v(q) = v_0 + \frac{g^2}{16\pi} \ln \frac {q_0}q\,, 
\end{equation}
%Since $\lambda\ll1$ the leading order result is
%\begin{equation}\label{n-weak}
%  n(q) = -\lambda(q) =  -\frac{g^2 N}{16 v(q)}
%  = -\frac{g^2N}{16} \left( v_0 
%  +\frac{g^2}{16\pi} \ln \frac {q_0} q\right)^{-1}
%\end{equation}
and the screening charge density is
\begin{equation}\label{n-weak}
  n(r) 
%  = \frac1{2\pi r^2} \frac{g^2N}{16} \frac{\beta(v(q))}{v^2(q)}
%  \biggl|_{q=1/r} 
  = \frac Z{Nr^2}\left( \frac{g^2N}{32\pi} \right)^2 \left( v_0 
  +\frac{g^2}{16\pi} \ln q_0r \right)^{-2}.
\end{equation}
Notice that the result is proportional to the square of the small
coupling constant $\lambda=g^2N/32v$, although we have performed the
calculation to leading order in the coupling.  The reason is that for
the charge density $n(r)$ to be nonzero, it is necessary that the
coupling constant runs.  The density $n(r)$ therefore contains the
beta function $\beta(v)$, as seen in Eq.~(\ref{nr-beta}), and hence is
second order in the coupling constant.

In the $1/N$ expansion the beta function for $v(q)$ was computed in
Ref.~\onlinecite{son}:
\begin{equation}
  \beta(v) = \left\{ \begin{array}{ll}
  -\displaystyle{\frac{8v}{\pi^2N}\left(  
  \frac{\ln(\lambda+\sqrt{\lambda^2-1})}
  {\lambda\sqrt{\lambda^2-1}}
    + 1 - \frac\pi{2\lambda}\right)}, & \qquad \lambda>1,
  \rule[-15pt]{0pt}{15pt}\\
   -\displaystyle{\frac{8v}{\pi^2N}\left(
   \frac{\arccos\lambda}{\lambda\sqrt{1-\lambda^2}}
    + 1 -\frac\pi{2\lambda}\right)}, & \qquad \lambda<1.
  \end{array}\right.   
\end{equation}
The two expressions smoothly match each other at $\lambda=1$.

In is instructive to analyze two regimes where the RG equation can be
solved analytically. The first regime is $\lambda\ll1$ where the
result is the same as in Eq.~(\ref{n-weak}).  The second regime is the
strong-coupling regime $\lambda\gg1$.
This regime corresponds to a quantum critical point characterized by
a dynamic critical exponent $z$, whose value at large $N$ is~\cite{son}
\begin{equation}
  z = 1 - \frac8{\pi^2N} + O(N^{-2}).
\end{equation}
In this regime $\beta=(z-1)v$.  The solution to the RG equation,
with the initial condition $v=v_0$ at $q=q_0$, is
\begin{equation}
  v(q) = v_0 \left( \frac{q_0} q\right)^{1-z}, \qquad
  1-z \approx \frac8{\pi^2N}\,.
\end{equation}
In this regime
\begin{equation}
  n(r) = \frac Z{2\pi r^2} \frac{1-z}{\lambda_0} (q_0 r)^{1-z}, 
  \qquad \lambda_0 = \frac{g^2N}{32v_0},
\end{equation}
i.e., the charge density follows a power law behavior $n(r)\sim
r^{-1-z}$.  The power is slightly different from $-2$.

\begin{figure}
   \begin{picture}(0,0)(0,0)
   \put(150,-10){$\log_{10}(r/r_0)$}
   \put(-50,100){$\displaystyle{\frac{2\pi}Z} r^2 n(r)$}
   \end{picture}
   \includegraphics[width=4in]{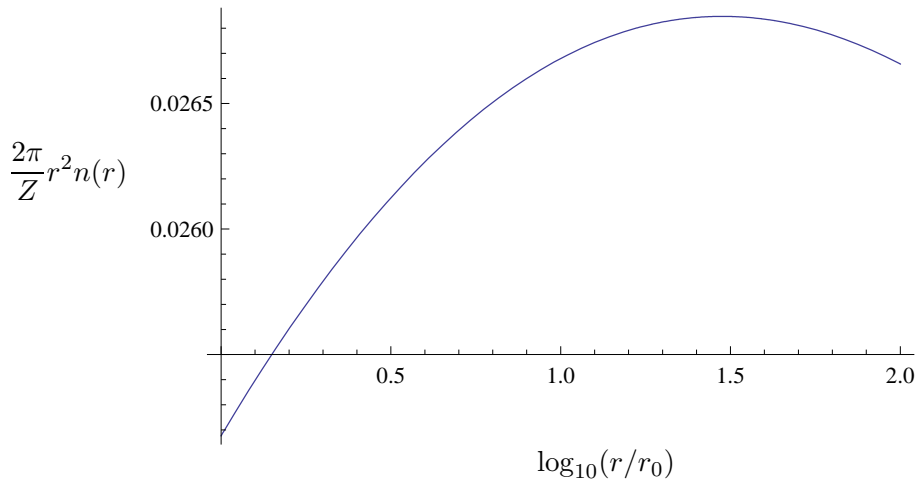}
\caption{The dependence of $2\pi Z^{-1} r^2n(r)$ on the distance $r$ for 
suspended graphene.  Note that coordinate $r$ is on a logarithmic scale.} 
\label{fig:scrcharge_vac}
\end{figure}

\begin{figure}
   \begin{picture}(0,0)(0,0)
   \put(150,-10){$\log_{10}(r/r_0)$}
   \put(-50,100){$\displaystyle{\frac{2\pi}Z} r^2 n(r)$}
   \end{picture}
   \includegraphics[width=4in]{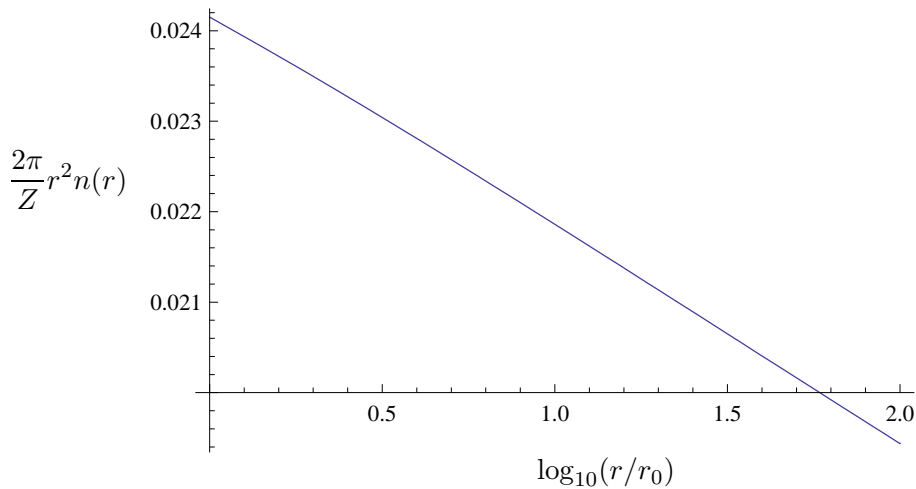}
\caption{The dependence of $2\pi Z^{-1} r^2 n(r)$ on the distance $r$ for
graphene on a substrate with $\epsilon=4.5$.  Note that
coordinate $r$ is on a logarithmic scale.} 
\label{fig:scrcharge_sub}
\end{figure}

In real graphene $\lambda$ is of order 1, so one
has to solve numerically the RG equation.  We chose the scale
$q_0$ to be comparable to the inverse lattice spacing, $r_0^{-1}$,
and $v_0$ to be
$10^6\textrm{m}/\textrm{s}$, a typical value found in experiments.  We
then run $v$ according to the leading (in $1/N$) RG equation in two cases, 
in vacuum and
when graphene is on a SiO$_2$ substrate with dielectric constant
$\epsilon=4.5$.  We then plot $2\pi r^2 n(r)$ as a function of the
distance $r$ on Figs.~(\ref{fig:scrcharge_vac}) and
(\ref{fig:scrcharge_sub}).

As seen from the figures, the charge density $n(r)$ roughly follows
the $r^{-2}$ law: when $r$ changes by two orders of magnitude, the
product $r^2 n(r)$ changes by a factor of less than 1.5 in both cases.

\section{Conclusions}
\label{sec:conc}

In this paper we have considered the problem of screening of 
a Coulomb impurity in graphene.
We show that there is a qualitative difference between screening by
non-interacting and interacting electrons.
In the case of non-interacting electrons the induced charge density is
localized at the position of the impurity when the impurity charge is small.
The interaction between electrons lead to a long-distance tail in
the induced charge distribution, with a counterintuitive sign which is
the same as that of the impurity.

One problem that is not addressed in this paper is the screening of an
impurity with large $\alpha\sim1$ by an interacting electron gas.
We hope to address this problem in a future publication.

\acknowledgments
An earlier version of this paper had a sign error in the $\alpha^3$ term in 
Eq.~(\ref{Falpha}); we thank V.~Kotov for pointing this out to us, and for giving us a preview of the work of Terekhov {\em et al.} \cite{kotov} which contains
a closed form expression for the function $F(\alpha)$.
The authors thank A.~V.~Andreev, M.~I.~Katsnelson, V.~N.~Kotov, and 
L.~S.~Levitov for useful discussions.
D.T.S. thanks the Center for Theoretical Physics at MIT, 
where part of this work was completed, for hospitality.
This work was supported, in part, by DOE Grant No.\
DE-FG02-00ER41132 and NSF Grant No.\ DMR-0537077.

\end{document}